\documentclass[
notoc,nohyper]{JHEP3} 

\newcommand{\Gammait}{{\mit\Gamma}}

\newcommand{\tr}{\mathop{\rm tr}\nolimits}

\newcommand{\Tr}{\mathop{\rm Tr}\nolimits}

\newcommand{\SU}{\mathop{\rm SU}}

\newcommand{\U}{\mathop{\rm {}U}}

\newcommand{\rmd}{{\rm d}}
\newcommand{\rmD}{{\rm D}}

\newcommand\fverb{\setbox\pippobox=\hbox\bgroup\verb}
\newcommand\fverbdo{\egroup\medskip\noindent%
                        \fbox{\unhbox\pippobox}\ }
\newcommand\fverbit{\egroup\item[\fbox{\unhbox\pippobox}]}
\newbox\pippobox

\title{
Anomalous gauge theories revisited}

\author{
Kosuke Matsui\\
Graduate School of Science and Engineering, Ibaraki University, Mito 310-8512,
Japan\\
E-mail: \email{matsui@serra.sci.ibaraki.ac.jp}}
\author{Hiroshi Suzuki\\
Institute of Applied Beam Science, Ibaraki University, Mito 310-8512, Japan\\
E-mail: \email{hsuzuki@mx.ibaraki.ac.jp}}

\received{\today}               
\accepted{\today}               

\preprint{IU-MSTP/66\\\heplat{0412041}}     

\abstract{
A possible formulation of chiral gauge theories with an anomalous fermion
content is re-examined in light of the lattice framework based on the
Ginsparg-Wilson relation. It is shown that the fermion sector of a wide class
of anomalous non-abelian theories cannot consistently be formulated within this
lattice framework. In particular, in 4~dimension, {\it all\/} anomalous
non-abelian theories are included in this class. Anomalous abelian chiral gauge
theories cannot be formulated with compact $\U(1)$ link variables, while a
non-compact formulation is possible at least for the vacuum sector in the space
of lattice gauge fields. Our conclusion is not applied to effective low-energy
theories with an anomalous fermion content which are obtained from an
underlying anomaly-free theory by sending the mass of some of fermions to
infinity. For theories with an anomalous fermion content in which the anomaly
is cancelled by the Green-Schwarz mechanism, a possibility of a consistent
lattice formulation is not clear.}


\begin{document} 

\maketitle 

\section{Introduction}
There had emerged a general belief that anomaly-free chiral gauge theories can
non-perturbatively be formulated, after a work by
L\"uscher~\cite{Luscher:1999du} which successfully formulated anomaly-free
abelian chiral gauge theories in a gauge invariant manner on a lattice.
Although there remain several challenging problems to be solved for a
non-abelian extension of this work, the problems are well-posed and the general
framework~\cite{Luscher:1999du,Luscher:1999un}, that is based on the so-called
Ginsparg-Wilson relation~\cite{Ginsparg:1982bj}, is ingenious. In fact, there
already appeared several theoretical applications of this framework for general
chiral gauge theories~\cite{Bar:2000qa,Luscher:2000zd,Fujiwara:2003np}.
Refs.~\cite{Kaplan:1992bt}--\cite{Golterman:2000hr} are related works for this
development.

The target of the lattice framework~\cite{Luscher:1999du,Luscher:1999un} is
anomaly-free chiral gauge theories. On the other hand, in the context of
continuum theory, it has sometimes been claimed that a consistent quantization
of chiral gauge theories is possible even when the gauge anomaly is not
cancelled~\cite{Jackiw:1984zi,Harada:1986wb,Miyake:1991ce}. The purpose of the
present paper is to clarify a situation concerning these {\it anomalous gauge
theories}, if one considers its quantization in the non-perturbative lattice
framework of refs.~\cite{Luscher:1999du,Luscher:1999un}. Actually, most of
expressions below (except those in the second half of section~5) has already
been known in the literature. However, since they are scattered in various
places where a quantization of anomalous gauge theories is not the main
subject, it seems useful to gather them together and summarize the situation in
a compact form.

In short, our general conclusion is the following: Generally, chiral gauge
theories with an anomalous fermion content cannot consistently be formulated
along a line of the framework of refs.~\cite{Luscher:1999du,Luscher:1999un}.
This conclusion is not related to (at least to our present understanding) a
loss of unitarity and/or renormalizability that is naively expected for gauge
theories with anomalies. Our conclusion is not even related to, at least
apparently, a loss of gauge invariance. Our point is that the partition
function of the fermion sector or more generally expectation values in the
fermion sector cannot be defined as a single-valued smooth function of the
gauge fields, if one applies the
framework~\cite{Luscher:1999du,Luscher:1999un} to theories with an anomalous
fermion content. This kind of situation does not happen in the continuum theory
in which a formal integration over fermion variables always defines a
single-valued smooth function of the gauge fields. This difference from the
continuum theory arises because in the framework
of~refs.~\cite{Luscher:1999du,Luscher:1999un}, one has to be involved with a
$\U(1)$ bundle associated to the fermion integration measure. A variation of
the partition function is related to the $\U(1)$ connection of this bundle,
called the measure term. There exists a wide freedom to choose the measure
term corresponding to a freedom to choose local counter terms. However, if the
$\U(1)$ bundle turns out to be non-trivial, the partition function cannot be a
single-valued smooth function of the gauge fields for any choice of the measure
term. These points will be clarified in the next section.

More specifically, for non-abelian gauge theories, we can show the breaking of
the smoothness occurs if $\pi_1(G)=0$ and $\pi_{2n+1}(G)=\mathbb{Z}$, where
$G$ is the gauge group and $2n$ is the dimension of the euclidean space, and a
fermion content exhibits an ``irreducible'' gauge anomaly.\footnote{The
irreducible here means that the anomaly cannot be expressed as a polynomial of
strictly lower rank traces over the fundamental representation of the gauge
group.} These conditions are precisely identical to prerequisites for a
geometrical characterization of the gauge anomaly studied in
ref.~\cite{Alvarez-Gaume:1983cs}; the phase of the chiral determinant possesses
a non-trivial winding number along a gauge loop and this implies a
non-contractible 2~sphere in the gauge orbit space~$\mathcal{A}/\mathcal{G}$
($\mathcal{A}$ denotes the space of gauge potentials and $\mathcal{G}$ the
group of gauge transformations). These prerequisites are applied to all
non-abelian anomalous gauge theories in 4~dimension so we conclude that
{\it all\/} 4~dimensional non-abelian anomalous gauge theories cannot
consistently be formulated with the present lattice framework~(section~3).

For abelian theories, on the other hand, the situation depends on whether one
uses compact link variables or non-compact variables for $\U(1)$ gauge fields.
For the compact $\U(1)$ case, the partition function in anomalous theories
cannot be a single-valued smooth function~(section~4). On the other hand, for
the non-compact $\U(1)$ case, one can arrange a suitable measure term to define
a physically acceptable fermion partition function at least for the vacuum
sector in the space of lattice gauge fields~(section~5). This is not surprising
because, with non-compact variables, the space of gauge fields (that is the
base space of the $\U(1)$ bundle) is topologically trivial and, consequently,
the $\U(1)$ bundle cannot be non-trivial.

We should emphasize that our conclusion is not applied to effective low-energy
theories with an anomalous fermion content which are obtained by sending the
mass of some of fermions infinity~\cite{D'Hoker:1984ph} from an underlying
anomaly-free theory. This and related issues will be discussed in section~6.

The dimension of the euclidean space will be denoted as~$d=2n$. Lorentz
indices, $\mu$, $\nu$, \dots\ run from 0 to~$d-1$. The hyper-cubic euclidean
lattice (whose one-dimensional size is~$L$) will be denoted by
$\Gammait=\{x\in\mathbb{Z}^d\mid0\leq x_\mu<L\}$. The unit vector along the
direction~$\mu$ will be denoted by~$\hat\mu$. The lattice spacing is taken to
be unity $a=1$ unless stated otherwise. The symbol
$\epsilon_{\mu_1\mu_2\cdots\mu_d}$ stands for the totally anti-symmetric
tensor with $\epsilon_{01\cdots d-1}=1$. We define the chiral matrix by
$\gamma_5=i^{-n}\gamma_0\gamma_1\cdots\gamma_{d-1}$ from hermitian Dirac
matrices~$\gamma_\mu$. The forward and backward difference operators on a
lattice are defined by $\partial_\mu f(x)=\{f(x+a\hat\mu)-f(x)\}/a$
and~$\partial_\mu^*f(x)=\{f(x)-f(x-a\hat\mu)\}/a$, respectively.

\section{Brief review of the lattice framework for Weyl
fermions~\cite{Luscher:1999du,Luscher:1999un}}
Our objective is the partition function, i.e., the Weyl determinant, of a
(left-handed) Weyl fermion on a lattice:\footnote{Our description will be
rather brief. For more details, readers are referred to
refs.~\cite{Luscher:1999du,Luscher:1999un}.} 
\begin{equation}
   \det M[U]
   =\int\rmD[\psi]\rmD[\overline\psi]\,e^{-S_{\rm F}},\qquad
   S_{\rm F}=\sum_{x\in\Gammait}\overline\psi(x)D\psi(x),
\label{twoxone}
\end{equation}
where $U$ denotes the lattice gauge field (link variables). A crucial
assumption on the lattice Dirac operator~$D$ is the Ginsparg-Wilson relation
\begin{equation}
   \gamma_5D+D\gamma_5=D\gamma_5D,
\label{twoxtwo}
\end{equation}
and the simplest example of such a lattice operator is provided by the
Neuberger overlap Dirac operator~\cite{Neuberger:1998fp}. The overlap Dirac
operator is free from the species doubling and possesses the
$\gamma_5$-hermiticity ($D^\dagger=\gamma_5D\gamma_5$), the gauge covariance
and moreover the locality\footnote{For a precise definition of the locality
assumed here, see ref.~\cite{Luscher:1999du}.} provided that configurations of
gauge fields are restricted by the admissibility~\cite{Hernandez:1999et}
\begin{equation}
   \|1-R[U(p)]\|<\epsilon,\qquad\hbox{for all plaquettes~$p$},
\label{twoxthree}
\end{equation}
where $U(p)$ is the product of link variables around the plaquette~$p$;
$R$~stands for a (generally reducible) unitary representation of the gauge
group to which the Weyl fermion belongs ($\epsilon$ is a certain fixed
constant). This condition divides otherwise arc-wise connected space of lattice
gauge fields into disconnected topological sectors.

The Ginsparg-Wilson relation allows one to introduce the modified chiral
matrix~$\hat\gamma_5=\gamma_5(1-D)$ which fulfills
\begin{equation}
   (\hat\gamma_5)^\dagger=\hat\gamma_5,\qquad(\hat\gamma_5)^2=1,\qquad
   D\hat\gamma_5=-\gamma_5D.
\label{twoxfour}
\end{equation}
Therefore, especially due to the last relation, the chirality of the Weyl
fermion can consistently be defined in the lattice action~$S_{\rm F}$ by
imposing the constraints
\begin{equation}
   \hat P_-\psi=\psi,\qquad\overline\psi=\overline\psi P_+,
\label{twoxfive}
\end{equation}
where we have introduced projection operators by
$\hat P_\pm=(1\pm\hat\gamma_5)/2$ and $P_\pm=(1\pm\gamma_5)/2$. Note that the
first constraint depends on lattice gauge fields through the Dirac
operator~$D$. The classical lattice action~$S_{\rm F}$ for the Weyl fermion is
then gauge invariant and local due to properties of~$D$ listed above.

We have to next define the fermion integration
measure~$\rmD[\psi]\rmD[\overline\psi]$ in eq.~(\ref{twoxone}). For this, we
introduce an orthonormal complete set of vectors in the constrained
space~(\ref{twoxfive}):
\begin{equation}
   \hat P_-v_j=v_j,\qquad (v_k,v_j)=\delta_{kj},
\label{twoxsix}
\end{equation}
and expand the fermion field~$\psi$ in terms of this basis as
$\psi(x)=\sum_jv_j(x)c_j$. The integration measure is then defined by
$\rmD[\psi]=\prod_j\rmd c_j$.\footnote{Similarly, the
anti-fermion~$\overline\psi$ is expanded by a basis such that
$\overline v_kP_+=\overline v_k$ as
$\overline\psi(x)=\sum_k\overline c_k\overline v_k(x)$ and the integration
measure is defined by $\rmD[\overline\psi]=\prod_k\rmd\overline c_k$.} In terms
of these bases, the Weyl determinant~(\ref{twoxone}) is given by the
determinant of the matrix~$M_{kj}=\sum_{x\in\Gammait}\overline v_k(x)Dv_j(x)$.

The above construction, although seemingly simple, turns out to be rather
involved due to following facts. First, the constraint~(\ref{twoxsix}) does not
specify the basis vectors uniquely. A different choice of basis leads to a
difference in the phase of the fermion integration measure and this phase
ambiguity may influence on physical contents of the system because the phase
may depend on gauge field configurations. One has to fix this phase ambiguity
so that basic physical requirements (smoothness, locality etc.) are fulfilled.
Secondly, a connected component of the space of admissible gauge fields,
denoted by $\mathfrak{U}$, may have a non-trivial topological structure. Hence
it is not obvious if there exists a basis over~$\mathfrak{U}$ with which above
physical requirements are fulfilled. These problems can be formulated as
follows~\cite{Luscher:1999du,Luscher:1999un}.

We cover~$\mathfrak{U}$ by local coordinate patches~$X_A$, where $A$ stands for
a label of patches. Suppose that a certain basis~$v_j^A$ has been chosen within
each of patch. Generally, however, on the intersection $X_A\cap X_B$ of two
patches, basis vectors $v_j^A$ and~$v_j^B$ are different; in general, these two
are related by a unitary transformation
\begin{equation}
   v_j^B(x)=\sum_k v_k^A(x)\tau(A\to B)_{kj}.
\label{twoxseven}
\end{equation}
By definition, the transition function~$\tau(A\to B)$ satisfies the cocycle
condition and thus it defines a unitary principal bundle over~$\mathfrak{U}$.
Corresponding to the relation~(\ref{twoxseven}), the fermion integration
measures defined in~$X_A$ and in~$X_B$ are related as
\begin{equation}
   \rmD[\psi]_B=\rmD[\psi]_Ag_{AB},\qquad g_{AB}=\det\tau(A\to B)\in\U(1).
\label{twoxeight}
\end{equation}
This phase factor~$g_{AB}$ thus defines a $\U(1)$ bundle associated to the
fermion integration measure. For the present formulation of the Weyl fermion to
be meaningful, the partition function~(\ref{twoxone}) (or more generally
expectation values in the fermion sector) must be a single-valued smooth
function over~$\mathfrak{U}$. To realize this, we re-define basis vectors in
each patch\footnote{Under a change of basis, the phase factor transforms as
$g_{AB}\to h_Ag_{AB}h_B^{-1}$, where $h_A$ is the determinant of the
transformation matrix between two bases in~$X_A$.} such that $g_{AB}=1$ on all
intersections. However, {\it this is possible if and only if the $\U(1)$ bundle
is trivial}.

To find a characterization of this $\U(1)$ bundle, we consider a variation of
gauge field
\begin{equation}
   \delta_\eta U(x,\mu)=\eta_\mu(x)U(x,\mu),\qquad
   \eta_\mu(x)=\eta_\mu^a(x)T^a,
\label{twoxnine}
\end{equation}
and define the ``measure term'' within a patch, say $X_A$, by
\begin{equation}
   \mathfrak{L}_\eta^A=i\sum_j(v_j^A,\delta_\eta v_j^A).
\label{twoxten}
\end{equation}
From the definition of the Weyl determinant, we have
\begin{equation}
   \delta_\eta\ln\det M=\Tr\{\delta_\eta D\hat P_-D^{-1}P_+\}
   -i\mathfrak{L}_\eta^A,
\label{twoxeleven}
\end{equation}
and this shows that for the Weyl determinant to be a single-valued smooth
function over~$\mathfrak{U}$, the measure term~$\mathfrak{L}_\eta$ must be
smoothly defined over~$\mathfrak{U}$. On the intersection, $X_A\cap X_B$, we
see from eq.~(\ref{twoxseven}) that the measure terms in $X_A$ and in~$X_B$ are
related as
\begin{equation}
   \mathfrak{L}_\eta^B=\mathfrak{L}_\eta^A+ig_{AB}^{-1}\delta_\eta
   g_{AB}.
\label{twoxtwelve}
\end{equation}
This shows that the measure term is the $\U(1)$ connection associated to the
$\U(1)$ bundle. Applying the identity
$\delta_\eta\delta_\zeta-\delta_\zeta\delta_\eta+\delta_{[\eta,\zeta]}=0$ to
the definition of the measure term~(\ref{twoxten}), we find\footnote{Here we
have assumed that the variations $\eta$ and $\zeta$ are independent of the
gauge field.} 
\begin{equation}
   \delta_\eta\mathfrak{L}_\zeta^A-\delta_\zeta\mathfrak{L}_\eta^A
   +\mathfrak{L}_{[\eta,\zeta]}^A
   =i\Tr\{\hat P_-[\delta_\eta\hat P_-,\delta_\zeta\hat P_-]\}.
\label{twoxthirteen}
\end{equation}
The right hand side is nothing but the curvature of the $\U(1)$ bundle. Thus
we can define the first Chern number of the $\U(1)$ bundle:
\begin{equation}
   \mathcal{I}={1\over2\pi}\int_{\mathcal{M}}\rmd t\,\rmd s\,
   i\Tr\{\hat P_-[\delta_\eta\hat P_-,\delta_\zeta\hat P_-]\},
\label{twoxfourteen}
\end{equation}
which is an {\it integer}. $\mathcal{M}$ in this expression stands for a
{\it closed\/} 2~dimensional surface in~$\mathfrak{U}$ and $t$ and~$s$ are
local coordinates of~$\mathcal{M}$. Namely, we defined 2~parameter family of
gauge fields $U_{t,s}$ in~$\mathfrak{U}$, and accordingly introduced the
projection operators by~$P_{t,s}=\hat P_-|_{U=U_{t,s}}$. The variations are
given by $\eta_\mu(x)=\partial_tU_{t,s}(x,\mu)U_{t,s}(x,\mu)^{-1}$ and
$\zeta_\mu(x)=\partial_sU_{t,s}(x,\mu)U_{t,s}(x,\mu)^{-1}$. The important point
for discussions in this paper is that if the above integer~$\mathcal{I}$ does
not vanish, then the $\U(1)$ bundle is non-trivial on~$\mathcal{M}$ and there
exists {\it no\/} possible choice of basis vectors such that the partition
function (or expectation values in the fermion sector) is a single-valued
smooth function over~$\mathcal{M}$. Namely, the Weyl fermion cannot
consistently be formulated. In this case, $\mathcal{M}$ is a non-contractible
2~surface in~$\mathfrak{U}$ which gives rise to a topological obstruction in
defining a smooth fermion measure.

In the above argument, we started with basis vectors defined patch by patch.
This way of argument, however, is not convenient in constructing the fermion
measure which is consistent with the locality. The measure
term~$\mathfrak{L}_\eta$, being linear in the variation~$\eta$, can be
expressed as
\begin{equation}
   \mathfrak{L}_\eta=\sum_{x\in\Gammait}\eta_\mu^a(x)j_\mu^a(x),
\label{twoxfifteen}
\end{equation}
where $j_\mu^a(x)$ is referred to as the measure current. It can be
shown~\cite{Luscher:1999du} that the expected locality of the system is
guaranteed if the measure current is a local function of lattice gauge fields.
To ensure the correct physical contents of the formulation, therefore, it is
convenient to start with a certain current which is a local function of gauge
fields. Then according to the reconstruction
theorem~\cite{Luscher:1999du,Luscher:1999un}, if the local current satisfies
the following conditions, one can re-construct basis vectors which lead to a
smooth fermion measure that is consistent with the locality. (1)~The current is
a single-valued and smooth function of the gauge fields contained
in~$\mathfrak{U}$. (2)~The measure term~(\ref{twoxfifteen}) defined from the
current satisfies the global integrability
\begin{equation}
   \exp\Bigl\{i\int_0^1\rmd t\,\mathfrak{L}_\eta\Bigr\}
   =\det\{1-P_0+P_0Q_1\},
\label{twoxsixteen}
\end{equation}
along {\it any\/} closed loop in~$\mathfrak{U}$. In this expression, the
loop~$U_t$ ($U_1=U_0$) is parametrized by~$t\in[0,1]$ and the variation is
given by $\eta_\mu(x)=\partial_tU_t(x,\mu)U_t(x,\mu)^{-1}$. Projection
operators along the loop are defined by~$P_t=\hat P_-|_{U=U_t}$. The
operator~$Q_t$ is defined by the differential equation~$\partial_t
Q_t=[\partial_tP_t,P_t]Q_t$ and~$Q_0=1$. Thus a task to construct
the fermion measure is reduced to find an appropriate local
current~$j_\mu^a(x)$ which satisfies the above two conditions.\footnote{For a
gauge invariant formulation for anomaly-free chiral gauge theories, an
additional condition, the anomaly cancellation on the
lattice~\cite{Luscher:1999du,Luscher:1999un}, has also to be fulfilled.} For a
contractible loop in $\mathfrak{U}$, it can be shown that the global
integrability is equivalent to the local integrability
\begin{equation}
   \delta_\eta\mathfrak{L}_\zeta-\delta_\zeta\mathfrak{L}_\eta
   +\mathfrak{L}_{[\eta,\zeta]}
   =i\Tr\{\hat P_-[\delta_\eta\hat P_-,\delta_\zeta\hat P_-]\},
\label{twoxseventeen}
\end{equation}
that is nothing but eq.~(\ref{twoxthirteen}) without patch labels (because here
we started with a measure term globally defined over~$\mathfrak{U}$).

\section{Non-abelian anomalous theories}
In this section, we assume that the gauge group~$G$ is semi-simple.
For these non-abelian gauge theories, we arrange following 2~parameter family
of gauge field configurations for which we can
evaluate~$\mathcal{I}$~(\ref{twoxfourteen}) (in this case, $\mathcal{M}=S^2$).
First, we consider a 1~parameter family of gauge transformations defined
on~$S^{2n}$ and take a lattice transcription of those gauge transformations:
\begin{equation}
   g_t(x)\in G,\qquad x\in\Gammait,\qquad 0\leq t\leq1,\qquad g_0(x)=g_1(x)=1.
\label{threexone}
\end{equation}
From these lattice gauge transformations, we define a 2~parameter family of
gauge field configurations:\footnote{Group elements, except those in a
measure-zero set in~$G$, are uniquely parametrized as $g=\exp(\theta^aT^a)$ by
using group generators~$T^a$. Powers of an element~$g$ are then defined by
$[g]^s=\exp(s\theta^aT^a)$. Powers in~eq.~(\ref{threextwo}) can always be
unambiguously defined at sufficiently small lattice spacings, because the
inside of the powers behaves as $1+O(a)$ for small lattice spacings. On the
other hand, we assume that the 1~parameter family~(\ref{threexone}) has been
chosen so that powers in eq.~(\ref{threexthree}) can be defined in the
aforementioned way.}
\begin{equation}
   U_{t,s}(x,\mu)=[g_t^{-1}(x)g_t(x+\hat\mu)]^s,\qquad 0\leq t\leq1,\qquad
   0\leq s\leq1.
\label{threextwo}
\end{equation}
In the continuum limit, these configurations reduce to linear interpolations
between the trivial vacuum and its continuum gauge transformations defined
by~$g_t(x)$. These configurations satisfy the admissibility~(\ref{twoxthree})
if the lattice is fine enough, because the left hand side of
eq.~(\ref{twoxthree}) is bounded by an $O(a^2)$ quantity. This 2~parameter
family thus spans a 2~disk~$D^2$ in $\mathfrak{U}$. We also define another
2~parameter family of gauge field configurations:
\begin{equation}
   U_{t,s}(x,\mu)=[g_{1-t}^{-1}(x)]^s[g_{1-t}(x+\hat\mu)]^s,\qquad
   0\leq t\leq1,\qquad 0\leq s\leq1.
\label{threexthree}
\end{equation}
This 2~parameter family spans another 2~disk~$D^2$ again in
$\mathfrak{U}$, because these configurations are lattice gauge transformations
of~$U(x,\mu)=1$ and the admissibility is a gauge invariant condition trivially
satisfied by~$U(x,\mu)=1$.\footnote{These configurations, however, are a pure
lattice construction and do not necessarily have a continuum limit. We notice
that their continuum limit is not needed in the evaluation of
eq.~(\ref{threexfour})~\cite{Fujiwara:2003np}.} We then glue the above two
disks~(\ref{threextwo}) and~(\ref{threexthree}) together along the edges $s=1$
and form a 2~sphere $\mathcal{M}=S^2$ in~$\mathfrak{U}$.

The first Chern number~$\mathcal{I}$~(\ref{twoxfourteen}) is an integer. With
the above setting, we can evaluate this integer in the classical continuum
limit. The result is~\cite{Fujiwara:2003np} (see also ref.~\cite{Adams:1999um})
\begin{eqnarray}
   \mathcal{I}&=&{(-1)^ni^{n+1}n!\over(2\pi)^{n+1}(2n+1)!}
   \int_{S^1\times S^{2n}}\rmd^{2n+1}x\,
\nonumber\\
   &&\qquad\times
   \epsilon_{\mu_1\mu_2\cdots\mu_{2n+1}}
   \tr\{R(g_t^{-1}\partial_{\mu_1}g_t)R(g_t^{-1}\partial_{\mu_2}g_t)\cdots
   R(g_t^{-1}\partial_{\mu_{2n+1}}g_t)\}
\nonumber\\
   &=&A_{n+1}(R){(-1)^ni^{n+1}n!\over(2\pi)^{n+1}(2n+1)!}
   \int_{S^1\times S^{2n}}\rmd^{2n+1}x\,
\nonumber\\
   &&\qquad\qquad\times
   \epsilon_{\mu_1\mu_2\cdots\mu_{2n+1}}
   \tr\{(g_t^{-1}\partial_{\mu_1}g_t)(g_t^{-1}\partial_{\mu_2}g_t)\cdots
   (g_t^{-1}\partial_{\mu_{2n+1}}g_t)\},
\label{threexfour}
\end{eqnarray}
where we regarded~$t$ as an additional coordinate~$x_{2n}$. In the second line,
$A_{n+1}(R)$ is the leading anomaly coefficient in $2n$~dimensions, defined
by\footnote{For the equality of the first line of eq.~(\ref{threexfour}) to the
second line, see, for example, ref.~\cite{Elitzur:1984kr}.}
\begin{eqnarray}
   &&\epsilon_{\mu_1\nu_1\cdots\mu_{n+1}\nu_{n+1}}
   \tr\{R(F_{\mu_1\nu_1})\cdots R(F_{\mu_{n+1}\nu_{n+1}})\}
\nonumber\\
   &&=A_{n+1}(R)\epsilon_{\mu_1\nu_1\cdots\mu_{n+1}\nu_{n+1}}
   \tr\{F_{\mu_1\nu_1}\cdots F_{\mu_{n+1}\nu_{n+1}}\}+
   (\hbox{lower traces}),
\end{eqnarray}
where the trace in the right hand side is defined with respect to the
fundamental representation of the gauge group. Note that, in 4 dimension, the
non-abelian gauge anomaly is always ``irreducible'', i.e., the gauge anomaly
always implies $A_{n+1}(R)\neq0$, because $\tr T^a=0$ for non-abelian factors.
To discuss in which circumstance the above
integer~$\mathcal{I}$~(\ref{threexfour}) does not vanish, we assume that
$\pi_1(G)=0$ and~$\pi_{2n+1}(G)=\mathbb{Z}$. Then, the
integer~$\mathcal{I}$~(\ref{threexfour}) is $A_{n+1}(R)$ times the winding
number of the gauge transformation~$g_t(x)$ around the basic ($2n+1$)~sphere
in~$G$~\cite{Alvarez-Gaume:1983cs}. One can then pick up $g_t(x)$ on
$S^1\times S^{2n}$ for which $\mathcal{I}\neq0$ if
$A_{n+1}(R)\neq0$.\footnote{Take a topologically
non-trivial mapping from $S^{2n+1}$ to $G$ which maps a certain point
of~$S^{2n+1}$ to the identity. The mapping can be regarded as a mapping from a
hypercube~$B^{2n+1}=[0,1]^{2n+1}$ to $G$ which maps all of the
boundary~$\partial B^{2n+1}$ to the identity. We regard
$B^{2n+1}=[0,1]\times B^{2n}$ and identity the first factor~$[0,1]$ with the
parameter~$t$ in eq.~(\ref{threexone}). For each value of~$t$, the mapping
defined above maps all of the boundary $\partial B^{2n}$ to the identity. So
this provides a mapping~$g_t(x)$ from $S^1\times S^{2n}$ to $G$ which gives a
non-zero winding number.}

In summary, if $\pi_1(G)=0$, $\pi_{2n+1}(G)=\mathbb{Z}$ and the gauge
anomaly is irreducible, i.e., $A_{n+1}(R)\neq0$, then we can provide an
explicit example of a 2~dimensional surface~$\mathcal{M}$ in $\mathfrak{U}$ for
which the first Chern number~(\ref{twoxfourteen}) is non-zero,
$\mathcal{I}\neq0$. We can thus infer that at least for such cases a consistent
lattice formulation of non-abelian anomalous gauge theories along the line of
framework of refs.~\cite{Luscher:1999du,Luscher:1999un} is impossible, because
the partition function (or expectation values in general) of the fermion sector
cannot be a single-valued smooth function over~$\mathcal{M}$. Although the
above cases do not exhaust all possible non-abelian gauge theories with an
anomalous fermion content, those are wide enough for one to expect a consistent
formulation of non-abelian anomalous cases is probably impossible in general.
In particular, in 4~dimension, non-abelian anomalies appear only when the gauge
group~$G$ contains $\SU(n)$ ($n\geq3$) as the factor (other groups have
anomaly-free representations only). Since $\pi_1(\SU(n))=0$
and~$\pi_5(\SU(n))=\mathbb{Z}$ for $n\geq3$, we conclude that {\it all\/}
non-abelian anomalous theories in 4~dimension cannot consistently be
formulated with the present lattice formulation.

\section{Abelian theory with compact gauge variables}
The obstruction for the smooth fermion measure discussed in the preceding
section does not exist in abelian theories because the right hand side of
eq.~(\ref{threexfour}) vanishes in this case. However, with a use of compact
$\U(1)$ link variables, it has been pointed out that there exists a
2~torus~$\mathcal{M}=T^2$ in $\mathfrak{U}$ for which the first Chern
number~(\ref{twoxfourteen}) is non-zero for an anomalous fermion content. This
implies that anomalous abelian chiral gauge theories cannot consistently be
formulated within the present lattice framework if compact $\U(1)$ link
variables are used.

The 2~torus in $\mathfrak{U}$ which is parametrized by coordinates $t$ and~$s$
($0\leq t,s<2\pi$) is defined by
\begin{equation}
   U_{t,s}(x,\mu)=
   \exp\{i\delta_{\mu,0}\delta_{\tilde x_0,0}t
   +i\delta_{\mu,1}\delta_{\tilde x_1,0}s\}V_{[m]}(x,\mu),
\label{fourxone}
\end{equation}
where $\tilde x_\mu=x_\mu\bmod L$, $0\leq\tilde x_\mu<L$. The Wilson line
$W_\mu(x)=\prod_{n=0}^{L-1}U(x+n\hat\mu,\mu)$ at the origin of these
configurations is given by $W_0(0)=e^{it}$ and~$W_1(0)=e^{is}$. An important
feature of these configurations is thus a winding along directions of the
Wilson line in the space of lattice gauge fields. The factor~$V_{[m]}(x,\mu)$
in the above expression carries a constant field strength
$F_{\mu\nu}=(1/i)\ln P(x,\mu,\nu)=2\pi m_{\mu\nu}/L^2$, where $m_{\mu\nu}$ are
integers. For integers within a range $|{\rm e}m_{\mu\nu}|<\epsilon'L^2/2\pi$,%
\footnote{$\epsilon'=2\arcsin(\epsilon/2)$.} where ${\rm e}$ is the
$\U(1)$ charge of the Weyl fermion, the configurations fulfill the
admissibility~(\ref{twoxthree}). The explicit form of~$V_{[m]}(x,\mu)$ is given
by
\begin{equation}
   V_{[m]}(x,\mu)
   =\exp\Bigl\{-{2\pi i\over L^2}\Bigl[
   L\delta_{\tilde x_\mu,L-1}\sum_{\nu>\mu}m_{\mu\nu}\tilde x_\nu
   +\sum_{\nu<\mu}m_{\mu\nu}\tilde x_\nu
   \Bigr]\Bigr\}.
\label{fourxtwo}
\end{equation}

For the 2~parameter family of configurations~(\ref{fourxone}), it can be shown
that~\cite{Kikukawa:2002ms} (see also
refs.~\cite{Neuberger:1998xn,Adams:1999um} for related study)
\begin{equation}
   \mathcal{I}={(-1)^{n-1}{\rm e}^{n+1}\over2^{n-1}(n-1)!}
   \epsilon_{01\mu_1\nu_1\cdots\mu_{n-1}\nu_{n-1}}
   m_{\mu_1\nu_1}\cdots m_{\mu_{n-1}\nu_{n-1}},
\label{fourxthree}
\end{equation}
when $L$, a one-dimensional size of the lattice, is large enough compared to
the localization range of the Dirac operator. In particular,
$\mathcal{I}={\rm e}^2$ for $d=2$. For the right-handed Weyl fermion, we have
eq.~(\ref{fourxthree}) with a minus sign. The Chern number~(\ref{twoxfourteen})
is therefore non-vanishing unless the gauge anomaly is cancelled among flavors
of the Weyl fermion as $\sum_\alpha{\rm H}_\alpha{\rm e}_\alpha^{n+1}=0$, where
${\rm H}_\alpha$ and ${\rm e}_\alpha$ are the chirality (${\rm H}_\alpha=\pm1$)
and the $\U(1)$ charge of a flavor~$\alpha$, respectively.

In the present compact~$\U(1)$ formulation, therefore, chiral gauge theories
with an anomalous fermion content in general cannot consistently be formulated.
This is a rather strong statement.

\section{Abelian theory with non-compact gauge variables}
The discussion in the preceding section does not exclude a possibility of a
formulation based on {\it non-compact\/} $\U(1)$ gauge variables:
\begin{equation}
   -\infty<A_\mu(x)<+\infty,\qquad x\in\Gammait,\qquad
   A_\mu(x+L\hat\nu)=A_\mu(x),
\end{equation}
because with these variables, a winding along directions of the Wilson line is
topologically trivial (see below). The gauge transformation on the non-compact
variables is defined by
\begin{equation}
   A_\mu^\omega(x)=A_\mu(x)-\partial_\mu\omega(x).
\label{fivextwo}
\end{equation}

In this formulation, the gauge potentials~$A_\mu(x)$ are regarded as primary
dynamical variables and a gauge coupling of the lattice fermion is defined by
substituting $\U(1)$ link variables in the lattice Dirac operator~$D$ by
\begin{equation}
   R[U(x,\mu)]=e^{i{\rm e}A_\mu(x)}.
\end{equation}
In the present non-compact formulation, the $\U(1)$ charge~${\rm e}$ is not
necessarily an integer.

The admissibility~(\ref{twoxthree}) restricts possible configurations of the
gauge potentials and divides the configuration space into disconnected sectors.
Here, for simplicity, we consider the vacuum sector which is a connected
component of the space of admissible configurations containing the trivial
vacuum~$A_\mu(x)=0$. By defining the field strength by
\begin{equation}
   F_{\mu\nu}(x)=\partial_\mu A_\nu(x)-\partial_\nu A_\mu(x),
\end{equation}
the vacuum sector is characterized by the condition
\begin{equation}
   |{\rm e}F_{\mu\nu}(x)|<\epsilon'.
\label{fivexfive}
\end{equation}
The vacuum sector in this non-compact formulation is topologically trivial
because when $A_\mu(x)$ satisfies eq.~(\ref{fivexfive}), then $tA_\mu(x)$ with
$0\leq t\leq1$ does too. Namely, any configuration in the vacuum sector can
smoothly be deformed to the trivial one $A_\mu(x)=0$ and, in particular, no
non-trivial winding along directions of the Wilson line is possible. In what
follows, we show that it is actually possible to set up a lattice formulation
for anomalous $\U(1)$ chiral gauge theories with these non-compact gauge
variables. For a related work for a lattice with an infinite extent, see
ref.~\cite{Neuberger:2000wq}. 

We define the partition function of the whole system as follows
\begin{equation}
   \mathcal{Z}=\int{\rm D}[A_\mu]\,e^{-S_{\rm G}-S_{\rm counter}}
   \int{\rm D}[\psi]{\rm D}[\overline\psi]\,e^{-S_{\rm F}},
\label{fivexsix}
\end{equation}
where ${\rm D}[A_\mu]=\prod_x\rmd A_\mu(x)$ and, as the gauge
action~$S_{\rm G}$, we adopt the modified plaquette
action~\cite{Luscher:1999du}
\begin{equation}
   S_{\rm G}={1\over4g_0^2}\sum_{x\in\Gammait}\sum_{\mu,\nu}
   \mathcal{L}_{\mu\nu},
\end{equation}
where $g_0$ is the bare coupling and
\begin{equation}
   \mathcal{L}_{\mu\nu}=\cases{[F_{\mu\nu}(x)]^2
   \{1-[{\rm e}F_{\mu\nu}(x)]^2/\epsilon^{\prime2}\}^{-1}
   &if $|{\rm e}F_{\mu\nu}(x)|<\epsilon'$\cr
   \infty&otherwise.\cr}
\label{fivexeight}
\end{equation}
The action $S_{\rm G}$ is designed to enforce the
admissibility~(\ref{fivexfive}) dynamically.\footnote{When the $\U(1)$
gauge field is coupled to several Weyl fermions with different $\U(1)$ charges,
the $\U(1)$ charge with a maximal magnitude~$|{\rm e}|$ is used in this
expression.}

We next have to show that the fermion integration measure
${\rm D}[\psi]{\rm D}[\overline\psi]$ can be defined smoothly on the space of
admissible gauge fields, in a way being consistent with the locality. It is
achieved by taking
\begin{equation}
   \mathfrak{L}_\eta=i\int_0^1\rmd t\,
   \Tr\{\hat P_-[\partial_t\hat P_-,\delta_\eta\hat P_-]\}
   +\delta_\eta\sum_{x\in\Gammait}A_\mu(x)\int_0^1\rmd t\,k_\mu(x),
\label{fivexnine}
\end{equation}
as the measure term in eq.~(\ref{twoxfifteen}). In this expression, the
parameter~$t$ interpolates the gauge potential from $0$ to~$A_\mu(x)$,
$A_\mu^t(x)=tA_\mu(x)$. It is understood that the gauge potentials~$A_\mu(x)$
within the integration over~$t$ are~$A_\mu^t(x)$. The current~$k_\mu(x)$ is a
smooth local function of the gauge
potentials which will be explained below. It is then obvious that
$\mathfrak{L}_\eta$ is a smooth function of admissible gauge fields. This is
also consistent with the locality of the theory because the measure
term~$\mathfrak{L}_\eta$ consists of a lattice sum of local expressions of the
gauge potentials due to the locality of the Dirac operator~$D$. One can also
verify that the measure term satisfies the local integrability
condition~(\ref{twoxseventeen}). Since the space of admissible gauge fields is
simply-connected in the present non-compact formulation, these guarantee that
there exist basis vectors which lead a smooth fermion integration measure being
consistent with the locality. This can easily be shown by constructing the
partition function~(\ref{twoxone}) explicitly which corresponds to the above
choice of the measure term. That is given by
\begin{equation}
   \ln\det M[A]=\int_0^1\rmd t\,\Tr\{\partial_tD\hat P_-D^{-1}P_+\}
   -i\sum_{x\in\Gammait}A_\mu(x)\int_0^1\rmd t\,k_\mu(x).
\label{fivexten}
\end{equation}
From this expression, by using relations
$\gamma_5\delta D=-\delta D\hat\gamma_5-D\delta\hat\gamma_5$ and
$\hat\gamma_5\delta P_-=-\delta\hat P_-\hat\gamma_5$, one finds
eq.~(\ref{twoxeleven}) with the measure term~(\ref{fivexnine}). The above form
of the measure term and the corresponding effective action have been known in
the literature~\cite{Luscher:1999du,Suzuki:1999qw}. We conclude therefore that
the criticism to anomalous gauge theories in previous sections based on a
smoothness of the fermion measure is not applied to the present case of
non-compact $\U(1)$ variables.

The partition function~(\ref{fivexten}) is not invariant under the gauge
transformation~(\ref{fivextwo}). We will see that the main part of the breaking
is given by the ``covariant gauge anomaly'' defined by
\begin{equation}
   \mathcal{A}(x)=-{1\over2}\tr\{\gamma_5{\rm e}D(x,x)\}.
\end{equation}
A remarkable fact is that one can deduce the explicit form of~$\mathcal{A}(x)$
in terms of the gauge potentials even on a finite lattice (as far as $L$ is
large enough) by using the local cohomology argument on a
lattice~\cite{Luscher:1998kn}--\cite{Kadoh:2003ii}. Although the local
cohomology argument was originally developed for the compact $\U(1)$ theory, an
adaptation to the present non-compact $\U(1)$ case is rather straightforward.
The result is~\cite{Igarashi:2002zz}
\begin{eqnarray}
   \mathcal{A}(x)&=&{(-1)^n{\rm e}^{n+1}\over(4\pi)^nn!}
   \epsilon_{\mu_1\nu_1\cdots\mu_n\nu_n}F_{\mu_1\nu_1}(x)
   F_{\mu_2\nu_2}(x+\hat\mu_1+\hat\nu_1)\cdots
\nonumber\\
   &&\qquad\qquad\qquad\times
   F_{\mu_n\nu_n}(x+\hat\mu_1+\hat\nu_1+\cdots+\hat\mu_{n-1}+\hat\nu_{n-1})
   +\partial_\mu^*k_\mu(x),
\end{eqnarray}
where the current~$k_\mu(x)$ is a certain gauge invariant local expression of
the gauge potentials. From the definition~(\ref{fivexten}) and the gauge
covariance of the Dirac operator, $D(x,y)[tA^\omega]=%
e^{it{\rm e}\omega(x)}D(x,y)[tA]e^{-it{\rm e}\omega(y)}$, we find
\begin{eqnarray}
   &&\ln\det M[A^{-\omega}]
\nonumber\\
   &&=\ln\det M[A]
   -i\sum_{x\in\Gammait}\omega(x)\int_0^1\rmd t\,
   [\mathcal{A}(x)-\partial_\mu^*k_\mu(x)]
\nonumber\\
   &&=\ln\det M[A]
   +i{(-1)^{n+1}{\rm e}^{n+1}\over(4\pi)^n(n+1)!}
   \sum_{x\in\Gammait}\omega(x)
   \epsilon_{\mu_1\nu_1\cdots\mu_n\nu_n}F_{\mu_1\nu_1}(x)
   F_{\mu_2\nu_2}(x+\hat\mu_1+\hat\nu_1)\cdots
\nonumber\\
   &&\qquad\qquad\qquad\qquad\qquad\qquad\qquad\times
   F_{\mu_n\nu_n}(x+\hat\mu_1+\hat\nu_1+\cdots+\hat\mu_{n-1}+\hat\nu_{n-1}).
\label{fivexthirteen}
\end{eqnarray}
The last term is nothing but the Wess-Zumino action~\cite{Wess:1971yu} in this
$\U(1)$ gauge theory. The Wess-Zumino action takes a particularly simple form
with our choice of the measure term~(\ref{fivexnine}).

With the definition of the Weyl determinant~(\ref{fivexten}), we may carry out
a study of anomalous $\U(1)$ gauge theories along the line of, say,
ref.~\cite{Harada:1986wb}. In the full partition
function~$\mathcal{Z}$~(\ref{fivexsix}), we can separate the integration over
gauge degrees of freedom by using the Faddeev-Popov trick. Namely, we insert
unity
\begin{equation}
   \Delta[A]\int{\rm D}[\omega]\prod_{x\in\Gammait}
   \delta(\partial_\mu^*A_\mu^\omega(x))=1,
\end{equation}
into $\mathcal{Z}$ where $\Delta[A]$ is the Faddeev-Popov determinant. Here,
we imposed the Lorentz gauge on the lattice. In the functional integration over
the gauge transformation~${\rm D}[\omega]$, we understand that the zero-mode
of~$\omega$ is excluded by the condition~$\sum_{z\in\Gammait}\omega(z)=0$.
Even with this condition, there exists a ``translational invariant'' measure
of~$\omega$ such that ${\rm D}[\omega+\lambda]={\rm D}[\omega]$. With this
exclusion of the zero-mode, the Faddeev-Popov determinant becomes well-defined
because  for each $A_\mu(x)$ the equation $\partial_\mu^*A_\mu^\omega(x)=0$ has
a unique solution within this space of~$\omega$:
\begin{equation}
   \omega(x)=\sum_{y\in\Gammait}G_L(x-y)\partial_\mu^*A_\mu(y),
\end{equation}
where $G_L(z)$ is the Green function of the lattice laplacian
\begin{equation}
   \partial_\mu^*\partial_\mu G_L(z)=\delta_{\tilde z,0}-L^{-2},\qquad
   G_L(z+L\hat\mu)=G_L(z),\qquad\sum_{z\in\Gammait}G_L(z)=0.
\end{equation}
As in the continuum theory, one can then confirm that the Faddeev-Popov
determinant is gauge invariant~$\Delta[A^\omega]=\Delta[A]$. By using the gauge
invariance of $\Delta[A]$ and~$S_{\rm G}$, we then arrive at the expression
\begin{equation}
   \mathcal{Z}=\int{\rm D}[A_\mu]\,\Delta[A_\mu]\prod_{x\in\Gammait}
   \delta(\partial_\mu^*A_\mu(x))\,e^{-S_{\rm G}}
   \int{\rm D}[\omega]\,e^{-S_{\rm counter}[A^{-\omega}]}\det M[A^{-\omega}].
\end{equation}

In what follows, we restrict our attention to 2~dimensional case ($n=1$),
i.e., the chiral Schwinger model, for which the classical continuum limit of
the Weyl determinant is amenable to a simple study. Namely, it is not difficult
to find that the classical continuum limit of eq.~(\ref{fivexten}) is
\begin{eqnarray}
   &&\ln\det M[A]
\nonumber\\
   &&=-{{\rm e}^2\over8\pi}\int\rmd^2x\,A_\mu(x)
   \Bigl\{\delta_{\mu\nu}
   -(\delta_{\mu\alpha}+i\epsilon_{\mu\alpha})
   {\partial_\alpha\partial_\beta\over\square}
   (\delta_{\beta\nu}-i\epsilon_{\beta\nu})
   \Bigr\}A_\nu(x)+O(a).
\label{fivexeighteen}
\end{eqnarray}
We can arrive this expression without any calculation of lattice Feynman
integrals. Assuming that the rotational symmetry is restored in the continuum
limit, a general argument on fermion one-loop diagrams in 2~dimension (see,
for example the appendix of~ref.~\cite{Fujikawa:2003az}) tells that
regularization ambiguities can appear only in a coefficient of the local
term~$A_\mu^2(x)$ in~eq.~(\ref{fivexeighteen}). Then the fact that our Weyl
determinant gives rise to the (consistent) gauge anomaly of the
form~(\ref{fivexthirteen}) completely fixes the coefficient of the
term~$A_\mu^2(x)$ as eq.~(\ref{fivexeighteen}). In traditional arguments in
quantization of this anomalous chiral Schwinger model~\cite{Jackiw:1984zi},
this regularization ambiguity in the local term~$A_\mu^2(x)$ is fully
utilized~\cite{Banerjee:1986tf}. So to incorporate this point to our
formulation, we further set the lattice counter terms as
\begin{equation}
   S_{\rm counter}={{\rm e}^2\over8\pi}(b-1)\sum_{x\in\Gammait}A_\mu^2(x),
\end{equation}
where $b$ is a free parameter. This completes our construction of the chiral
Schwinger model on the lattice. It yields
\begin{eqnarray}
   \mathcal{Z}&=&\int{\rm D}[A_\mu]\,\prod_{x\in\Gammait}
   \delta(\partial_\mu^*A_\mu(x))\,e^{-S_{\rm G}-S_{\rm counter}[A]}
   \det M[A]
   \int{\rm D}[\omega]
\nonumber\\
   &&\times
   \exp\Bigl\{-{{\rm e}^2\over8\pi}\sum_{x\in\Gammait}\Bigl[
   (b-1)\partial_\mu\omega(x)\partial_\mu\omega(x)
   -2(b-1)\omega(x)\partial_\mu^*A_\mu(x)
   -i\omega(x)\epsilon_{\mu\nu}F_{\mu\nu}(x)\Bigr]\Bigr\},
\nonumber\\
\label{fivextwenty}
\end{eqnarray}
where we have got rid of the Faddeev-Popov determinant~$\Delta[A]$ because it
is a constant for the present Lorentz gauge. The action in
eq.~(\ref{fivextwenty}) except the gauge fixing term
\begin{eqnarray}
   &&S_{\rm G}[A]+S_{\rm counter}[A]-\ln\det M[A]
\nonumber\\
   &&+{{\rm e}^2\over8\pi}\sum_{x\in\Gammait}\Bigl[
   (b-1)\partial_\mu\omega(x)\partial_\mu\omega(x)
   -2(b-1)\omega(x)\partial_\mu^*A_\mu(x)
   -i\omega(x)\epsilon_{\mu\nu}F_{\mu\nu}(x)\Bigr],
\end{eqnarray}
is gauge invariant under $A_\mu(x)\to A_\mu(x)-\partial_\mu\lambda(x)$
and~$\omega(x)\to\omega(x)+\lambda(x)$ because it was
originally~$S_{\rm G}[A]+S_{\rm counter}[A^{-\omega}]-\ln\det M[A^{-\omega}]$
and the combination $A^{-\omega}$ is invariant under the transformation.
After integrating over~$\omega(x)$, we have
\begin{eqnarray}
   \mathcal{Z}&=&\int{\rm D}[A_\mu]\,\prod_{x\in\Gammait}
   \delta(\partial_\mu^*A_\mu(x))
\nonumber\\
   &&\times\exp\Bigl\{
   -{1\over4g_0^2}\sum_{x\in\Gammait}\sum_{\mu,\nu}\mathcal{L}_{\mu\nu}
   +{m^2\over4g_0^2}
   \sum_{x,y\in\Gammait}F_{\mu\nu}(x)G_L(x-y)F_{\mu\nu}(y)\Bigr\}
   \det M[A]
\nonumber\\
   &&\times
   \exp\Bigl\{{{\rm e}^2\over8\pi}\sum_{x,y\in\Gammait}
   A_\mu(x)\Bigl[\delta_{\mu\nu}\delta_{x,y}
   -(\partial_\mu^x+i\epsilon_{\mu\alpha}\partial_\alpha^{x*})
   G_L(x-y)(\partial_\nu^{y*}-i\partial_\beta^y\epsilon_{\beta\nu})\Bigr]
   A_\nu(y)\Bigr\}
\nonumber\\
   &&\times
   \exp\Bigl\{-{{\rm e}^2b\over8\pi L^2}
   \Bigl[\sum_{x\in\Gammait}A_\mu(x)\Bigr]^2\Bigr\},
\end{eqnarray}
where the mass parameter~$m^2$ is given
by~$m^2=({\rm e}^2g_0^2/4\pi)b^2/(b-1)$. In an intermediate stage in deriving
the above expression, we have used the identity~$\partial_\mu\partial_\nu^*=%
\delta_{\mu\nu}\partial_\rho^*\partial_\rho+%
\epsilon_{\mu\rho}\partial_\rho^*\partial_\sigma\epsilon_{\sigma\nu}$ holding
on a 2~dimensional lattice. It is easy to verify that the action in the above
expression is still gauge invariant. From eqs.~(\ref{fivexeight})
and~(\ref{fivexeighteen}), the classical continuum limit of the above action
is given by
\begin{equation}
   -{1\over4g_0^2}\int\rmd^2x\,F_{\mu\nu}(x)
   {\square-m^2\over\square}F_{\mu\nu}(x)
   =-{1\over2g_0^2}\int\rmd^2x\,A_\mu(x)
   (-\square+m^2)A_\mu(x),
\end{equation}
under the gauge condition $\partial_\mu A_\mu(x)=0$. This appearance of massive
excitations in the anomalous chiral Schwinger model is in accord with the
result of refs.~\cite{Jackiw:1984zi,Harada:1986wb}.

\section{Discussion}

In this paper, we showed that a consistent lattice formulation of anomalous
chiral gauge theories along the line of framework of
refs.~\cite{Luscher:1999du,Luscher:1999un} is rather difficult and for a
certain class of models it is in fact impossible. For an implication of our
observation, one can take two alternative attitudes. One may of course regard
obstructions for a smooth fermion measure we discussed as a pathology being
peculiar to the present lattice framework based on the Ginsparg-Wilson
relation. For example, it might be possible to formulate anomalous non-abelian
chiral theories by using certain non-compact gauge variables, as we have
demonstrated for the abelian case. With such non-compact gauge variables,
however, an issue of how a non-trivial topology of gauge field configurations
emerges on a lattice has to be clarified.

Alternatively, it is possible to take our observation rather seriously and
consider its possible implications. We may regard it as an indication for an
impossibility of a consistent quantization of anomalous chiral gauge theories
in general. (The successful treatment of abelian theories in section~5 is
regarded exceptional.) We emphasize that obstructions we observed do not exist
in effective low-energy theories with an anomalous fermion content, which are
obtained by sending mass of some of fermions very large (say, by sending the
expectation value of the Higgs field very large) from an underlying
anomaly-free theory in which the gauge anomaly is cancelled among flavors of
the Weyl fermion. From the way of construction in section~2, the effect of the
measure term is left over even if the mass of a Weyl fermion is sent to
infinity (like the Wess-Zumino term in the continuum
theory~\cite{D'Hoker:1984ph}).\footnote{The Yukawa interaction can be
introduced without affecting to the chirality constraint~(\ref{twoxfive}).}
Then the obstruction we discussed is cancelled among fermion flavors provided
that the gauge anomaly is cancelled among them. The effect of the measure term
does not decouple even the mass of fermion is very large.

This latter attitude is compatible with a general belief that anomalous gauge
theories are inconsistent anyway and the fundamental theory must be free
from anomalies after all. This is OK. However, a trouble with the latter
attitude is the fact that a cancellation among flavors of the Weyl fermion is
{\it not\/} an only known way for a cancellation of the gauge anomaly. The
Green-Schwarz mechanism~\cite{Green:1984sg} is another known way for the
anomaly cancellation. In this mechanism, the anomaly arising from Weyl fermions
is cancelled by bosonic anti-symmetric tensor fields by assuming a non-trivial
gauge transformation law of latter fields. Then the question is: can we set up
a non-perturbative lattice formulation of anomaly-free theories based on the
Green-Schwarz mechanism? In the context of the present lattice framework based
on the Ginsparg-Wilson relation, to answer this question in a positive way, we
have to find a certain interplay between the measure term of Weyl fermions and
a definition of anti-symmetric tensor fields on a lattice. Such an interplay
seems not straightforward to find at the moment. We have to admit, therefore,
the situation is not clear for chiral gauge theories with an anomalous fermion
content in which the anomaly {\it is\/} cancelled by the Green-Schwarz
mechanism.

H.S. would like to thank Kazuo Fujikawa for an enlightening discussion
concerning gauge theories with anomalies. This work is supported in part by
Grant-in-Aid for Scientific Research, \#13135203 and \#15540250.

\paragraph{Note added in proofs.}
The partition function of Weyl fermions in an anomalous multiplet can always be
made gauge invariant by introducing a dynamical Wess-Zumino scalar. This way of
anomaly cancellation can be regarded as the simplest form of the Green-Schwarz
mechanism. At least this form of the Green-Schwarz mechanism can readily be
implemented in lattice gauge theories by (1)~replacing link
variables~$U(x,\mu)$ by $U^g(x,\mu)=g(x)U(x,\mu)g(x+\hat\mu)^{-1}$ and
(2)~integrating over the Wess-Zumino scalar~$g$ with an appropriate gauge
invariant kinetic term. Under the gauge transformation,
$\delta U(x,\mu)=-\nabla_\mu\omega(x)U(x,\mu)$ and
$\delta g(x)=-g(x)\omega(x)$, the combination $U^g$ is gauge invariant and thus
the partition function of the fermion sector is trivially gauge invariant. We
then examine a smoothness of the fermion measure as a function of
configurations of the gauge field and of the Wess-Zumino scalar. The curvature
of the $\U(1)$ bundle associated to the fermion measure defined over this
``enlarged'' configuration space is given by the right hand side of
eq.~(\ref{twoxthirteen}) with substitutions $U\to U^g$,
$\eta_\mu\to g\eta_\mu g^{-1}$ and~$\zeta_\mu\to g\zeta_\mu g^{-1}$. Because of
the gauge covariance of the projection operator~$\hat P_-$, however, this
curvature turns out to have an identical form as in the case without the
Wess-Zumino scalar. Hence, the obstructions for a smooth fermion measure we
discussed in this paper remain in this enlarged theory, although the gauge
anomaly is exactly cancelled. We conclude that a large class of anomalous
chiral gauge theories, even with this simplest form of the Green-Schwarz
mechanism, cannot consistently be formulated in the present lattice framework.

\listoftables           
\listoffigures          


\begin{thebibliography}{99}

\bibitem{Luscher:1999du}
M.~L\"uscher,
\emph{Abelian chiral gauge theories on the lattice with exact gauge
invariance},
\npb{549}{1999}{295}
[\heplat{9811032}].

\bibitem{Luscher:1999un}
M.~L\"uscher,
\emph{Weyl fermions on the lattice and the non-abelian gauge anomaly},
\npb{568}{2000}{162} [\heplat{9904009}].

\bibitem{Ginsparg:1982bj}
P.~H.~Ginsparg and K.~G.~Wilson,
\emph{A remnant of chiral symmetry on the lattice},
\prd{25}{1982}{2649}.

\bibitem{Bar:2000qa}
O.~B\"ar and I.~Campos,
\emph{Global anomalies in chiral gauge theories on the lattice},
\npb{581}{2000}{499} [\heplat{0001025}].
\\
O.~B\"ar,
\emph{On Witten's global anomaly for higher $\SU(2)$ representations},
\npb{650}{2003}{522} [\heplat{0209098}].

\bibitem{Luscher:2000zd}
M.~L\"uscher,
\emph{Lattice regularization of chiral gauge theories to all orders of
perturbation theory},
\jhep{06}{2000}{028} [\heplat{0006014}].

\bibitem{Fujiwara:2003np}
T.~Fujiwara, K.~Matsui, H.~Suzuki and M.~Yamamoto,
\emph{Wess-Zumino-Witten term on the lattice},
\jhep{09}{2003}{015} [\heplat{0307031}].

\bibitem{Kaplan:1992bt}
D.~B.~Kaplan,
\emph{A method for simulating chiral fermions on the lattice},
\plb{288}{1992}{342} [\heplat{9206013}].

\bibitem{Narayanan:1993wx}
R.~Narayanan and H.~Neuberger,
\emph{Infinitely many regulator fields for chiral fermions},
\plb{302}{1993}{62} [\heplat{9212019}];
\emph{Chiral determinant as an overlap of two vacua},
\npb{412}{1994}{574} [\heplat{9307006}];
\emph{Chiral fermions on the lattice},
\prl{71}{1993}{3251} [\heplat{9308011}];
\emph{A construction of lattice chiral gauge theories},
\npb{443}{1995}{305} [\hepth{9411108}].

\bibitem{Randjbar-Daemi:1995sq}
S.~Randjbar-Daemi and J.~Strathdee,
\emph{On the overlap formulation of chiral gauge theory},
\plb{348}{1995}{543} [\hepth{9412165}];
\emph{Chiral fermions on the lattice},
\npb{443}{1995}{386} [\heplat{9501027}];
\emph{On the overlap prescription for lattice regularization of chiral
fermions},
\npb{466}{1996}{335} [\hepth{9512112}];
\emph{Consistent and covariant anomalies in the overlap formulation of
chiral gauge theories},
\plb{402}{1997}{134} [\hepth{9703092}].

\bibitem{Kikukawa:1997qh}
Y.~Kikukawa and H.~Neuberger,
\emph{Overlap in odd dimensions},
\npb{513}{1998}{735} [\heplat{9707016}].

\bibitem{Hasenfratz:1998ft}
P.~Hasenfratz,
\emph{Prospects for perfect actions},
\npps{63}{1998}{53} [\heplat{9709110}];
\emph{Lattice QCD without tuning, mixing and current renormalization}.
\npb{525}{1998}{401} [\heplat{9802007}].
\\
P.~Hasenfratz, V.~Laliena and F.~Niedermayer,
\emph{The index theorem in QCD with a finite cut-off},
\plb{427}{1998}{125} [\heplat{9801021}].

\bibitem{Luscher:1998pq}
M.~L\"uscher,
\emph{Exact chiral symmetry on the lattice and the Ginsparg-Wilson relation},
\plb{428}{1998}{342} [\heplat{9802011}].

\bibitem{Narayanan:1998uu}
R.~Narayanan,
\emph{Ginsparg-Wilson relation and the overlap formula},
\prd{58}{1998}{097501} [\heplat{9802018}].
\\
F.~Niedermayer,
\emph{Exact chiral symmetry, topological charge and related topics},
\npps{73}{1999}{105} [\heplat{9810026}].

\bibitem{Golterman:2000hr}
M.~Golterman,
\emph{Lattice chiral gauge theories},
\npps{94}{2001}{189} [\heplat{0011027}].
\\
Y.~Kikukawa,
\emph{Analytic progress on exact lattice chiral symmetry},
\npps{106}{2002}{71} [\heplat{0111035}].

\bibitem{Jackiw:1984zi}
R.~Jackiw and R.~Rajaraman,
\emph{Vector meson mass generation through chiral anomalies},
\prl{54}{1985}{1219} [Erratum-ibid.\ {\bf 54} (1985) 2060].
\\
L.~D.~Faddeev and S.~L.~Shatashvili,
\emph{Realization of the Schwinger term in the Gauss law and the possibility of
correct quantization of a theory with anomalies},
\plb{167}{1986}{225}.

\bibitem{Harada:1986wb}
K.~Harada and I.~Tsutsui,
\emph{On the path integral quantization of anomalous gauge theories},
\plb{183}{1987}{311}.
\\
O.~Babelon, F.~A.~Schaposnik and C.~M.~Viallet,
\emph{Quantization of gauge theories with Weyl fermions},
\plb{177}{1986}{385}.

\bibitem{Miyake:1991ce}
S.~Miyake and K.~Shizuya,
\emph{Consistency and current algebras of a non-Abelian chiral gauge theory in
four-dimensions}, \prd{45}{1992}{2090}, and references therein.

\bibitem{Alvarez-Gaume:1983cs}
L.~Alvarez-Gaum\'e and P.~H.~Ginsparg,
\emph{The topological meaning of nonabelian anomalies},
\npb{243}{1984}{449}.

\bibitem{D'Hoker:1984ph}
E.~D'Hoker and E.~Farhi,
\emph{Decoupling a fermion whose mass is generated by a Yukawa coupling: The
general case},
\npb{248}{1984}{59}.

\bibitem{Neuberger:1998fp}
H.~Neuberger,
\emph{Exactly massless quarks on the lattice},
\plb{417}{1998}{141} [\heplat{9707022}];
\emph{More about exactly massless quarks on the lattice},
\plb{427}{1998}{353} [\heplat{9801031}].

\bibitem{Hernandez:1999et}
P.~Hern\'andez, K.~Jansen and M.~L\"uscher,
\emph{Locality properties of Neuberger's lattice Dirac operator},
\npb{552}{1999}{363} [\heplat{9808010}].
\\
H.~Neuberger,
\emph{Bounds on the Wilson Dirac operator},
\prd{61}{2000}{085015} [\heplat{9911004}].

\bibitem{Adams:1999um}
D.~H.~Adams,
\emph{Index of a family of lattice Dirac operators and its relation to the
non-abelian anomaly on the lattice},
\prl{86}{2001}{200} [\heplat{9910036}];
\emph{Global obstructions to gauge invariance in chiral gauge theory on the
lattice},
\npb{589}{2000}{633} [\heplat{0004015}];
\emph{Families index theory for overlap lattice Dirac operator. I},
\npb{624}{2002}{469} [\heplat{0109019}];
\emph{Gauge fixing, families index theory, and topological features of the
space of lattice gauge fields},
\npb{640}{2002}{435} [\heplat{0203014}];
\emph{Fermionic topological charge of families of lattice gauge fields},
\npps{119}{2003}{775} [\heplat{0210052}].

\bibitem{Elitzur:1984kr}
S.~Elitzur and V.~P.~Nair,
\emph{Nonperturbative anomalies in higher dimensions},
\npb{243}{1984}{205}.

\bibitem{Kikukawa:2002ms}
Y.~Kikukawa and H.~Suzuki,
\emph{Chiral anomalies in the reduced model},
\jhep{09}{2002}{032} [\heplat{0207009}].

\bibitem{Neuberger:1998xn}
H.~Neuberger,
\emph{Geometrical aspects of chiral anomalies in the overlap},
\prd{59}{1999}{085006}
[\heplat{9802033}].

\bibitem{Neuberger:2000wq}
H.~Neuberger,
\emph{Noncompact chiral $\U(1)$ gauge theories on the lattice},
\prd{63}{2001}{014503} [\heplat{0002032}].

\bibitem{Suzuki:1999qw}
H.~Suzuki,
\emph{Gauge invariant effective action in Abelian chiral gauge theory on the
lattice},
\ptp{101}{1999}{1147} [\heplat{9901012}].

\bibitem{Luscher:1998kn}
M.~L\"uscher,
\emph{Topology and the axial anomaly in abelian lattice gauge theories},
\npb{538}{1999}{515} [\heplat{9808021}].

\bibitem{Fujiwara:1999fi}
T.~Fujiwara, H.~Suzuki and K.~Wu,
\emph{Non-commutative differential calculus and the axial anomaly in abelian
lattice gauge theories},
\npb{569}{2000}{643} [\heplat{9906015}];
\emph{Axial anomaly in lattice abelian gauge theory in arbitrary dimensions},
\plb{463}{1999}{63} [\heplat{9906016}].
\\
H.~Suzuki,
\emph{Anomaly cancellation condition in lattice gauge theory},
\npb{585}{2000}{471} [\heplat{0002009}].
\\
H.~Igarashi, K.~Okuyama and H.~Suzuki,
\emph{Errata and addenda to ``Anomaly cancellation condition in lattice gauge
theory}'', \heplat{0012018}.
\\
Y.~Kikukawa and Y.~Nakayama,
\emph{Gauge anomaly cancellations in $\SU(2)_L\times\U(1)_Y$ electroweak theory
on the lattice},
\npb{597}{2001}{519} [\heplat{0005015}].
\\
Y.~Kikukawa,
\emph{Domain wall fermion and chiral gauge theories on the lattice with exact
gauge invariance},
\prd{65}{2002}{074504} [\heplat{0105032}].

\bibitem{Igarashi:2002zz}
H.~Igarashi, K.~Okuyama and H.~Suzuki,
\emph{More about the axial anomaly on the lattice},
\npb{644}{2002}{383} [\heplat{0206003}].

\bibitem{Kadoh:2003ii}
D.~Kadoh, Y.~Kikukawa and Y.~Nakayama,
\emph{Solving the local cohomology problem in $\U(1)$ chiral gauge theories
within a finite lattice},
\jhep{12}{2004}{006} [\heplat{0309022}].
\\
D.~Kadoh and Y.~Kikukawa,
\emph{A numerical solution to the local cohomology problem in $\U(1)$ chiral
gauge theories}, \heplat{0401025}.

\bibitem{Wess:1971yu}
J.~Wess and B.~Zumino,
\emph{Consequences of anomalous Ward identities},
\plb{37}{1971}{95}.

\bibitem{Fujikawa:2003az}
K.~Fujikawa and H.~Suzuki,
\emph{Anomalies, local counter terms and bosonization},
\prep{398}{2004}{221} [\hepth{0305008}].

\bibitem{Banerjee:1986tf}
R.~Banerjee,
\emph{One parameter class of solutions in the chiral Schwinger model},
\prl{56}{1986}{1889}.

\bibitem{Green:1984sg}
M.~B.~Green and J.~H.~Schwarz,
\emph{Anomaly cancellation in supersymmetric $D=10$ gauge theory and
superstring theory},
\plb{149}{1984}{117}.

\end{thebibliography}
\end{document}